\documentclass[sigconf]{acmart}

\usepackage{booktabs}
\usepackage{graphicx}
\usepackage{tikz}
\usepackage{xspace}
\usetikzlibrary{arrows.meta,fit,positioning}

\newcommand{\hbfsim}{HBFSim\xspace}

\newcounter{takeaway}
\newcommand{\takeaway}[2]{\par\refstepcounter{takeaway}%
    \textbf{Takeaway~\thetakeaway: #1} #2\par}

\copyrightyear{2026}
\acmYear{2026}
\setcopyright{cc}
\setcctype{by-nc-nd}
\acmConference[RACS '26]{International Conference on Resilience, AI and Cyber-Systems}{November 17--20, 2026}{Fukuoka, Japan}
\acmBooktitle{International Conference on Resilience, AI and Cyber-Systems (RACS '26), November 17--20, 2026, Fukuoka, Japan}
\acmDOI{10.1145/3848591.3856870}
\acmISBN{979-8-4007-3035-1/2026/11}

\begin{document}

\title[HBFSim]{HBFSim: A Simulator for Exploring HBF in Memory Hierarchies for LLMs}

\author{Jinheng Li}
\orcid{0000-0003-2279-7973}
\affiliation{%
  \institution{Mohamed bin Zayed University of Artificial Intelligence}
  \city{Abu Dhabi}
  \country{United Arab Emirates}
}
\email{jinheng.li@mbzuai.ac.ae}

\author{Guosheng Wang}
\orcid{0009-0005-5062-641X}
\affiliation{%
  \institution{Mohamed bin Zayed University of Artificial Intelligence}
  \city{Abu Dhabi}
  \country{United Arab Emirates}
}
\email{guosheng.wang@mbzuai.ac.ae}

\author{Hongchao Du}
\orcid{0000-0001-7995-1834}
\affiliation{%
  \institution{Mohamed bin Zayed University of Artificial Intelligence}
  \city{Abu Dhabi}
  \country{United Arab Emirates}
}
\email{hongchao.du@mbzuai.ac.ae}

\author{Shangyu Wu}
\orcid{0000-0002-1961-143X}
\affiliation{%
  \institution{Mohamed bin Zayed University of Artificial Intelligence}
  \city{Abu Dhabi}
  \country{United Arab Emirates}
}
\email{shangyu.wu@mbzuai.ac.ae}

\author{Qiao Li}
\orcid{0000-0002-4579-4268}
\affiliation{%
  \institution{Mohamed bin Zayed University of Artificial Intelligence}
  \city{Abu Dhabi}
  \country{United Arab Emirates}
}
\email{qiao.li@mbzuai.ac.ae}

\author{Chun Jason Xue}
\orcid{0000-0002-6431-9868}
\affiliation{%
  \institution{Mohamed bin Zayed University of Artificial Intelligence}
  \city{Abu Dhabi}
  \country{United Arab Emirates}
}
\email{jason.xue@mbzuai.ac.ae}

\renewcommand{\shortauthors}{Li et al.}

\begin{abstract}
High Bandwidth Flash (HBF) has recently emerged as a promising technology that provides terabyte-scale accelerator-local capacity at bandwidths approaching HBM, with a potential cost-per-bit advantage over HBM. 
However, its NAND foundation brings long latency, out-of-place updates, and finite write endurance directly into the accelerator memory hierarchy, making its effects on performance and lifetime difficult to predict.  
To study this emerging design space, we present \hbfsim, an LLM-aware event-driven simulator for heterogeneous HBM+HBF memory systems that preserves application-level memory semantics while modeling flash-internal behavior. \hbfsim models HBM, HBF with host-managed flash translation, and conventional capacity tiers in one simulation driven by a calibrated serving frontend. 
Using \hbfsim, we study decoding of Llama-3.1-70B and Qwen3-235B-A22B under three workloads that differ in their input and output lengths. 
HBF serves data that does not fit in HBM 11.0--19.0$\times$ faster than host DRAM; HBM is better used to hold the running sequences' KV cache than as a cache for HBF; and at a published peak load of Llama-3.1-70B, HBF lasts 2.0 to 6.9 years, in proportion to the stacks that share the KV writes. The artifact is available at \url{https://github.com/FlashLabAE/HBFSim}
and will track public HBF device data as it becomes available.
\end{abstract}



\makeatletter
\def\@ACM@copyright@check@cc{}
\makeatother

\maketitle

\section{Introduction}
Memory capacity has emerged as a primary bottleneck in serving large language models (LLMs). Model weights occupy tens to hundreds of gigabytes, while key--value (KV) state grows linearly with context length and request concurrency, quickly exceeding accelerator memory capacity. For example, an 8B model serving eight concurrent 128K-context sessions can exhaust the HBM of a modern GPU (e.g., the 131 GiB of HBM3e on an NVIDIA H200), while long-context serving for 70B-class models may exceed the memory capacity of an entire eight-GPU node~\cite{kwon_pagedattention,qin_mooncake,dubey_llama3,nvidia_h200}. High Bandwidth Flash (HBF) has recently emerged as a promising approach for extending accelerator-local memory capacity. SanDisk's disclosed HBF design targets a 16-die, 512 GB flash stack that delivers up to 1.6 TB/s~\cite{sandisk_hbf_fact_sheet,sandisk_hbf_blog}, enabling multi-terabyte accelerator-local memory systems without host-memory accesses. The same disclosure targets 8--16$\times$ HBM capacity at similar cost and bandwidth, making capacity per dollar, rather than higher speed, the main motivation for HBF~\cite{sandisk_hbf_fact_sheet}. 

Unlike conventional DRAM expansion, HBF brings asymmetric read and write performance, page programming, flash translation layer (FTL) management, write amplification, and finite program/erase endurance into the accelerator memory hierarchy. Its effect on LLM serving is difficult to predict because no public HBF hardware is available for validation~\cite{skhynix_hbf_standardization}. Nevertheless, system architects must decide on memory organization, placement policies, and provisioning before production devices become available.
For example, moving read-only weights to HBF mainly exercises its read path and frees HBM for the growing KV cache, whereas placing mutable KV state in HBF creates program traffic and consumes write endurance. 
Therefore, it is necessary to study the performance of HBM+HBF memory hierarchies under realistic LLM serving workloads before committing to a design. Because such hardware is unavailable, a simulator is required to evaluate these trade-offs.

Existing simulators each cover one level of this hierarchy: DRAM simulators~\cite{luo_ramulator2,li_dramsim3} stop at the memory controller, SSD simulators~\cite{mqsim,femu} put flash behind a storage interface, and LLM serving simulators~\cite{zhang_llmcompass,gtsim,khairy_accelsim} represent memory by capacity and bandwidth. None connects the placement of LLM data to both completion time and physical flash traffic.

\hbfsim addresses this limitation. It models HBM, HBF with a host-managed FTL, and conventional capacity tiers in one event system, coupled in a closed loop with a serving frontend whose scheduler and GPU compute are calibrated against measurements, and it accepts address-level traces as well as production request traces.
Using \hbfsim, we study decoding of Llama-3.1-70B and Qwen3-235B-A22B under three workloads that differ in their input and output lengths. We find that HBF serves the data that does not fit in HBM 11.0--19.0$\times$ faster than host DRAM; that HBM should hold the running sequences' KV, because caching HBF with LRU or FIFO rereads nearly the whole working set once it exceeds HBM, and that KV should be written to HBF only when it must leave HBM; and that the FTL costs little on the read path as long as its layout keeps reads parallel, while HBF's lifetime depends on the amount of KV that reaches flash. These findings connect HBF's potential cost-per-bit advantage to measurable latency and wear rather than to a single capacity or bandwidth number.

% Figure 1 (cited in Section 2.1) is declared here so that it floats to the top of this column.
\begin{figure}[t]
    \centering
    \includegraphics[width=\columnwidth]{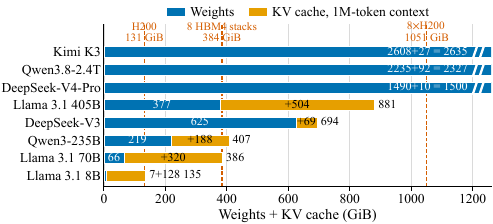}
    \vspace{-8pt}
    \caption{Weights (one byte per parameter) and BF16 KV cache of a
    1M-token context against the HBM of one H200, eight HBM4 stacks, and an
    eight-H200 node.}
    \Description{Stacked horizontal bars of weights and KV cache in
    gibibytes: Llama 3.1 8B 7 plus 128, Llama 3.1 70B 66 plus 320,
    Qwen3-235B 219 plus 188, DeepSeek-V3 625 plus 69, Llama 3.1 405B 377 plus
    504, and, cut at the end of the axis, DeepSeek-V4-Pro 1490 plus 10,
    Qwen3.8-2.4T 2235 plus 92, and Kimi K3 2608 plus 27. Dashed vertical lines
    mark 131 gibibytes for an H200, 384 for eight HBM4 stacks, and 1,051 for
    eight H200s.}
    \label{fig:motivation}
    \vspace{-10pt}
\end{figure}

This paper makes the following contributions:
\begin{itemize}
    \item \textbf{Workloads at two granularities.} \hbfsim takes fine-grained address-level traces, with one record per memory access, and coarse-grained request traces, which its frontend expands into the series of weight and KV accesses that each request causes.

    \item \textbf{A coupled and fast memory-system model.} HBM, HBF with a host-managed FTL, and external tiers (host DRAM, CXL SSDs, and NVMe SSDs) share one timeline and interact as in a real system, for example through the FTL's tables in HBM.

    \item \textbf{Preliminary takeaways for HBF.} Decoding Llama-3.1-70B and Qwen3-235B-A22B under three workloads yields five simple rules: HBF pays off only under capacity pressure; HBM should hold the running KV and HBF the weights, with KV written to flash only when it must leave HBM; generic cache replacement loses to this placement unless a step fits in HBM with room to spare; the FTL should preserve the flash array's parallelism; and endurance, set by how much KV reaches flash, is a first-order constraint.

\end{itemize}

\section{Background}
\label{sec:bg}

\subsection{Memory Demand of LLM Serving}

A serving system's memory holds the model weights, which every request reuses; the KV cache, the per-layer keys and values of every processed token, which decoding reads instead of recomputing; and activations, which are small by comparison. 
For Llama-3.1-70B a BF16 KV cache takes 320\,KiB per token~\cite{dubey_llama3}, 320\,GiB for 1M tokens. Serving systems allocate KV in fixed-size blocks~\cite{kwon_pagedattention} and retain the blocks of finished requests so that later requests with the same prefix skip its prefill~\cite{zheng_sglang,qin_mooncake}, so the KV footprint grows beyond the active requests. Figure~\ref{fig:motivation} shows the weights and the KV cache of 1M tokens, in one request or retained across many: for small models the KV cache dominates, every model from 70B up exceeds eight HBM4 stacks, and the newest open models, with 1.6T to 2.8T parameters, exceed even an eight-H200 node with their weights alone, while hybrid or compressed attention keeps their KV small.

The two phases of a request access this data differently. Prefill processes the prompt, produces its KV, and is usually compute-bound; chunked prefill lets it share a batch with decoding work~\cite{agrawal_sarathi}. Each decoding step generates one token per request: it reuses the weights across the batch, reads every running request's KV, and appends one token's KV per layer, so it is usually bound by memory bandwidth, and its traffic grows with model size, batch, and context. When weights and KV exceed HBM, systems offload to host DRAM or SSDs~\cite{sheng_flexgen,liu_retrievalattention,qin_mooncake}, at access costs set by the slower tier and its interconnect.

\vspace{-6pt}
\subsection{High Bandwidth Flash}

HBF stacks NAND flash dies on a logic base die and connects them to the
accelerator through an HBM-like interface in the same
package~\cite{sandisk_hbf_fact_sheet,sandisk_hbf_blog}. 
The OCP HBF specification~\cite{ocp_hbf} defines stacks with up to 16 channels and three speed grades.
For example, at speed grade~2 each channel carries 96\,GB/s of payload, or about 1.5\,TB/s per
stack, and SanDisk targets 512\,GB per stack. An HBF stack therefore offers
8-16 times the capacity of an HBM stack at a similar interface
bandwidth.

However, HBF still behaves like NAND. A page read takes microseconds, compared to tens of nanoseconds of DRAM, so HBF reaches its bandwidth only with many
reads in flight. Data is programmed, not overwritten: a page program takes
tens of microseconds and occupies its bank, and a block must be erased before
its pages are programmed again. The specification assigns address mapping,
garbage collection (GC), and wear leveling to host software; the device
itself performs neither GC nor copyback~\cite{ocp_hbf}. Each block tolerates
a limited number of program/erase (P/E) cycles, and no P/E rating for HBF has
been published. Host decisions thus shape both performance and lifetime:
what the host places in HBF determines the traffic HBF sees, and the host's
flash translation layer (FTL) determines how many erases each write costs.

\vspace{-6pt}
\subsection{Questions for HBM+HBF Systems}
\label{sec:questions}

Before HBF hardware is available, system designers need to answer five
questions. Is HBF a better capacity tier than host DRAM or SSDs? How much HBM
should accompany it? Which data should stay in HBM, and should HBM act as a
cache for HBF or as separately addressed memory? How should the FTL handle
KV writes? How long will HBF last under serving traffic? The answers
depend on each other through the serving loop: slower memory stretches
decoding steps, which changes how requests are batched and therefore the
memory traffic itself. Answering them requires running a real serving
scheduler against memory models that include flash management, with the
two coupled in a closed loop.

\section{\hbfsim Design}
\label{sec:design}

\begin{figure}[t]
    \centering
    \resizebox{\columnwidth}{!}{\begin{tikzpicture}[
    x=1cm,y=1cm,
    >={Latex[length=1.6mm]},
    flow/.style={->,semithick,draw=black!75},
    group/.style={draw=black!45,rounded corners=1.5pt,line width=0.4pt},
    block/.style={draw=black!55,rounded corners=1.2pt,align=center,
                  inner xsep=2pt,inner ysep=2pt,
                  font=\fontsize{6.5}{7.5}\selectfont,
                  execute at begin node={\hyphenpenalty=10000\exhyphenpenalty=10000}},
    front/.style={block,fill=gray!10,text width=1.82cm,minimum height=0.9cm},
    policy/.style={block,fill=purple!10,text width=3.85cm,minimum height=0.8cm},
    hbm/.style={block,fill=blue!11,text width=1.5cm,minimum height=1.05cm},
    ftl/.style={block,fill=orange!28,text width=5.15cm,minimum height=0.45cm},
    hbf/.style={block,fill=orange!13,text width=3.4cm,minimum height=1.05cm},
    tier/.style={block,fill=green!11,text width=2.3cm,minimum height=1.05cm},
    grouptitle/.style={anchor=west,inner sep=0pt,font=\footnotesize\bfseries},
    note/.style={inner sep=1pt,font=\fontsize{6.5}{7.5}\selectfont,fill=white},
    loop/.style={->,semithick,draw=red!70!black},
    couple/.style={->,thick,draw=purple!70!black},
]

% Serving frontend
\draw[group,fill=gray!3] (0,4.46) rectangle (8.2,5.98);
\node[grouptitle] at (0.14,5.8) {Serving frontend};
\node[front] (trace) at (0.98,5.0) {Request traces\\arrivals, lengths,\\prefix hashes};
\node[front] (sched) at (3.05,5.0) {Scheduler\\batching, paged\\KV, prefix cache};
\node[front] (comp)  at (5.12,5.0) {Compiler\\layer DAG: reads,\\writes, prefetch};
\node[front] (gpu)   at (7.19,5.0) {GPU timing\\calibrated kernels,\\memory overlap};
\draw[flow] (trace) -- (sched);
\draw[flow] (sched) -- (comp);
\draw[flow] (comp) -- (gpu);
\draw[loop] (gpu.north) -- (7.19,5.65) -- node[note,above,pos=0.5,text=red!70!black,fill=none] {batch time} (3.05,5.65) -- (sched.north);

% Memory policies
\draw[group,fill=purple!3] (0,2.88) rectangle (8.2,4.08);
\node[grouptitle] at (0.14,3.9) {Memory policies};
\node[note,anchor=east,fill=none] at (8.06,3.9) {objects $\rightarrow$ device addresses};
\node[policy] (peer)  at (2.1,3.3) {Peer mode\\each object in either memory; HBM\\first to the weights or the running KV};
\node[policy] (cache) at (6.1,3.3) {Tiered mode\\HBM caches HBF in 4\,MiB lines:\\LRU, FIFO, decayed LFU; write-back};
\draw[flow] (comp.south) -- node[note,right=1pt,pos=0.45] {object accesses} (5.12,4.08);

% Device engine
\draw[group,fill=blue!2] (0,0.26) rectangle (8.2,2.5);
\node[grouptitle] at (0.14,2.32) {Device engine};
\node[note,anchor=east,fill=none] at (8.06,2.33) {one timeline, per-resource calendars};
\node[ftl]  (ftl)  at (2.775,1.92) {Host FTL: mapping, GC, wear leveling};
\node[hbm]  (hbm)  at (0.95,0.9) {HBM\\channels,\\arbitration};
\node[hbf]  (hbf)  at (3.65,0.9) {HBF device\\channels, planes, page buffers;\\read, program, erase timing;\\device DRAM buffer and cache};
\node[tier] (ext)  at (6.75,0.9) {Other tiers\\host DRAM, CXL SSD,\\NVMe SSD: link,\\controller, media};
\draw[flow] (3.65,1.695) -- (hbf.north);
\draw[couple] (0.95,1.695) -- (hbm.north);
\node[note,anchor=west,fill=none,text=purple!70!black] at (1.05,1.56) {tables, GC copies};
\draw[flow] (4.10,2.88) -- node[note,right=1pt,pos=0.5] {transactions with dependencies} (4.10,2.5);

% Closed loop and results
\draw[loop] (8.2,0.9) -- (8.56,0.9) -- (8.56,5.0)
    node[pos=0.5,sloped,below,font=\fontsize{6.5}{7.5}\selectfont,text=red!70!black] {memory feedback}
    -- (gpu.east);
\node[font=\fontsize{6.5}{7.5}\selectfont,align=center,anchor=north] at (4.10,0.2)
    {results: time per output token, bytes per object class and device, programs and erases per block};
\end{tikzpicture}}
    \vspace{-8pt}
    \caption{\hbfsim structure: three layers. Feedback from the device
    engine closes the loop to the scheduler, and the host FTL keeps its tables and GC copies
    in HBM.}
    \Description{Block diagram with three layers. The serving frontend
    contains request traces, a scheduler, a compiler that emits each
    layer's reads, writes and prefetches with dependencies, and a GPU
    timing model; an arrow labeled batch time returns from GPU timing to
    the scheduler. Memory policies map objects to device addresses in peer
    mode or tiered mode. The device engine runs one timeline with
    per-resource calendars and contains HBM, a host flash translation layer
    above the HBF device and HBM, and other capacity tiers; an arrow labeled
    tables and GC copies goes from the FTL into HBM, and an arrow labeled
    memory feedback goes from the device engine back to GPU timing.}
    \label{fig:architecture}
    \vspace{-10pt}
\end{figure}

\hbfsim is a closed-loop simulator of LLM serving on accelerator memory
built from HBM, HBF, and other capacity tiers. Figure~\ref{fig:architecture}
shows its three layers. A serving frontend turns request traces into batches
and the requests in each batch into memory accesses. 
A policy layer decides in which memory each
object is stored and how data moves between tiers. A device engine executes the memory transactions on models of HBM, HBF, or external capacity
tiers, and feeds the outcome of that work back to the frontend, which schedules the next batch based on the feedback. 
Slower memory therefore reshapes batching as it would on a real system: longer iterations let more requests queue before the next one starts, and when KV no longer fits, the scheduler moves some KV to a slower tier or preempts requests, which shrinks the batch.
Three challenges shape this design.

\textbf{Workloads and device accesses differ in granularity and semantics.} A
serving system works with requests, weights, and KV blocks; HBM serves data bursts
of tens of bytes; and HBF is accessed like memory but programs flash pages (e.g., 4\,KiB) in
order, erases whole flash blocks, and leaves its FTL to the host. Address-level
traces preserve individual accesses but can only cover short runs, whereas production
request traces cover long periods of serving but record no memory accesses.

\textbf{The levels of the memory hierarchy are coupled.} Existing simulators model one level in isolation, but in an HBM+HBF system the two levels jointly determine performance. What HBF must serve depends on what HBM can hold, and the traffic HBF generates in turn consumes HBM capacity and bandwidth. Placement decisions therefore shift load between the tiers rather than simply adding a slower level beneath HBM, so neither tier can be simulated independently.

\textbf{The number of events is huge.} One decoding step of a 70B model reads
tens of millions of flash pages, each passing through about a dozen shared resources, from the plane and flash channel to the TSVs and ECC engine.
Because slower memory changes batching, these accesses must be simulated in a
closed loop rather than replayed.

As shown in Figure~\ref{fig:architecture}, layers in \hbfsim communicate through narrow interfaces. The frontend describes work
in terms of model objects such as weight tensors and KV blocks. Policies map objects to devices and emit transaction
graphs with explicit dependencies. The engine contains no LLM-specific logic. This
separation lets users change one layer, for example, add a caching policy or
an FTL, without touching the others.

\vspace{-6pt}
\subsection{Serving Frontend}
\label{sec:frontend}

\hbfsim accepts workloads at two granularities. 
In an \emph{address-level trace}, a record is
one memory access with optional data class, arrival time, and layer
annotations; such traces are replayed directly. 
Detailed instruction-and-address tracing and cycle-level simulation incur substantial costs even for a small inference workload. For Qwen2.5-1.5B-Instruct (bfloat16, batch size 1), a 32-token prefill followed by two fixed-token decode steps required 43.4 minutes for trace collection on an RTX 4000 Ada with NVBit~\cite{villa_nvbit}. The resulting 1,030-kernel traces occupy 98.3 GiB uncompressed, or 7.46 GiB in the compressed format consumed by Accel-Sim2.0~\cite{accelsim2_2026}. Full trace-driven simulation already takes 19.9 wall-clock hours even for a small model. For larger models, both the trace size and simulation time would become prohibitively large, rendering this approach unrealistic.

In a \emph{request-level trace}, a record is one inference request: its arrival time,
prompt and output lengths, and prefix hashes~\cite{qin_mooncake,wang_kvcache_wild}.
It contains no detailed memory accesses.
Which bytes an
inference step touches does not depend on the values it computes, only on
the model architecture and on which requests the scheduler runs with how
many tokens (for MoE models, also on the routed experts, which depend on
the values and are therefore an input, recorded or synthetic). The frontend, ServeLoop, therefore regenerates the accesses of every
step. Its scheduler replays the trace with vLLM's continuous batching and
chunked prefill~\cite{yu_orca,kwon_pagedattention,agrawal_sarathi},
including paged KV, prefix sharing, and preemption. A compiler turns each
iteration into the byte ranges that every operator reads or writes in weight
tensors, KV blocks, and embedding rows, leaving addresses and memories to
the policies. A decoding step of Llama-3.1-70B under the workload with 32 sequences and 6K tokens of context becomes $1.8\times10^{5}$ such transactions.

The generated accesses are verified at three levels. Batching: given the
same service times, the scheduler forms the same batches as vLLM
(Section~\ref{sec:validation}). Addresses: the byte ranges and their order
match 6,568 held-out warp streams captured with NVBit on an A100; the H200
we model has no capture, so its KV is read block by block, which changes the
time per output token (TPOT) by at most 0.55\% against the A100 order.
Volume: the frontend counts each byte once, whereas prefill kernels refetch
the same KV for other query heads or tiles, which GPU caches partly absorb,
so prefill traffic is a lower bound.

\hbfsim times every batch, prefill and decode alike, using an operator-level GPU model. It decomposes each batch into operators and computes each operator's time from three components: a fixed overhead, a compute time, and a memory-access time. The compute time is derived from the operator's FLOPs and a compute-throughput coefficient fitted to measured kernels. We obtain the memory-access time from \hbfsim's detailed simulation of HBM and HBF accesses, rather than an ideal-bandwidth assumption. To stay anchored to real GPU behavior, we fit each operator's fixed overhead, compute throughput, and memory bandwidth from measured kernels to establish a baseline time. During simulation, we replace the baseline's ideal memory time with the simulated access time, while retaining the measured kernel's additional memory overhead beyond the GPU's peak bandwidth. The batch time is then the sum of all operator times. This design ensures that HBF latency and bandwidth directly affect each operator's memory time, and thus the batch time, while the compute portion stays tied to measured GPU performance.

\vspace{-6pt}
\subsection{Memory Policies}
\label{sec:policies}

\hbfsim provides two ways to organize HBM and HBF. In \emph{peer mode}, HBM and HBF form one address space, and each object is stored in one of them. HBM is filled in a priority order and the rest overflows to HBF: \emph{weights first} places the weights, then KV blocks in address order, retained prefixes included; \emph{KV first} places the running sequences' KV, then the weights, then the retained prefixes. In \emph{tiered mode}, HBM caches objects stored in the HBF or other capacity
tier in 4\,MiB lines with LRU, FIFO, or decayed-LFU~\cite{robinson_fbr} replacement; decayed
LFU evicts the line with the fewest recent accesses and halves all access
counts periodically, as in frequency-based replacement~\cite{robinson_fbr}.
Dirty data is tracked at 4\,KiB granularity, and dirty KV is written back
either eagerly, while compute leaves the link idle~\cite{lee_eager_writeback},
or only when its line is evicted.
KV released before eviction is discarded without being
written. 
% A third policy prefetches the next layer's weights into bounded HBM
% buffers while the current layer computes. Every data movement is issued as a
% transaction with its dependencies, so copies compete for the same resources
% as demand accesses.

\vspace{-6pt}
\subsection{Device Engine}
\label{sec:device}

\textbf{HBM.} An HBM stack follows the JEDEC organization of its
generation, set by the number of channels and pseudo-channels and the pin
rate. The following study uses HBM4 for analysis~\cite{jedec_hbm4}, with 32 channels of two
pseudo-channels at 8\,Gb/s per pin, or 2\,TB/s per stack. Because our studies concern capacity tiers, \hbfsim models HBM at the
pseudo-channel level rather than at the command level of DRAM simulators such
as Ramulator~2.0~\cite{luo_ramulator2}. Requests are split into pseudo-channel
transfers, groups of four pseudo-channels share an arbitration queue and a
service timeline, and an average access latency and a bandwidth efficiency
account for row activation and refresh.

\textbf{HBF.} As in the OCP specification, the model separates the device
from its host-side management. The device has independent transmit and
receive timelines for each channel, ordered page sensing within each bank,
page buffers, ECC, and TSV transfers. It programs the pages of a block in
order and, as the specification requires, erases a block when its first
page is written. The organization of flash into channels, dies, planes, and
blocks and the timing of reads, programs, and erases follow
MQSim~\cite{mqsim}. We reimplemented them because MQSim's FTL runs inside
an SSD, whereas HBF's runs on the host, and HBF operations must share one
timeline with HBM and the serving frontend. Like an SSD, and beyond the
specification, each stack has a device-managed DRAM that buffers writes
until a page is full and caches recently read pages.
The host runs the FTL. It maps logical pages to physical ones and runs
garbage collection, which picks victim blocks by their number of valid pages
with a small weight for wear; static wear leveling is implemented but
switched off in our studies. Since the specification forbids the device
from moving valid data, garbage collection reads live pages into a copy
buffer in HBM and writes them back. We implemented three mappings and
keep each one's index in HBM. Page-level mapping has an 8-byte entry per
4\,KiB page; the whole table lives in flash as translation pages, and HBM
caches part of it. Extent mapping keeps a sorted list of runs of
consecutive pages, 48\,B per run. Both write by appending and spread
consecutive pages over all planes. Block-level mapping has one entry per
1\,MiB logical block and keeps the block in one physical block on one plane,
each page at a fixed offset. We model two consequences: NAND programs a
block's pages in order, so a page whose predecessors have not been written
yet waits in the device DRAM; and any write that is not an in-order append
copies the whole block to a fresh one. We do not stripe blocks over planes,
because a copy would then move a 1.5\,GiB stripe. Every mapping gets 0.1\%
of the HBF capacity in HBM, 3.1\,GiB of the 96\,GiB of two HBM4 stacks, and
its lookups compete with the application's traffic: with page-level
mapping, one Llama-3.1-70B decoding step at 2~HBM + 6~HBF issues
1.0--1.7$\times10^{7}$ lookups, one per flash page read.

\textbf{Other tiers.} Host DRAM, CXL SSDs, and NVMe SSDs share one
model consisting of a link, a controller, and media queues. A CXL SSD adds
a device-side DRAM cache with FIFO replacement.

\vspace{-6pt}
\subsection{Simulation Speed}
\label{sec:speed}
\begin{figure}[t]
    \centering
    \includegraphics[width=\columnwidth]{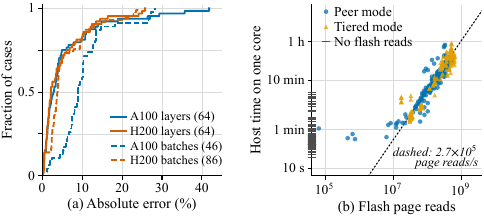}
    \vspace{-8pt}
    \caption{Calibration and simulation cost. (a) Absolute error of
    predicted time for held-out layer shapes and for complete-request
    batches. (b) Host time of the 405 configurations of
    Section~\ref{sec:cases} against their flash page reads; configurations
    without flash reads sit on the vertical axis.}
    \Description{Two panels. The first shows four cumulative
    distributions of absolute prediction error: 64 held-out layers on the
    A100 and on the H200, 46 batches on the A100, and 86 batches on the
    H200; most errors are below 10 percent and the largest is 41.8 percent.
    The second is a scatter plot in which host time grows in proportion to
    flash page reads, from about one minute at ten million reads to just
    under an hour at six hundred million.}
    \label{fig:validation}
    \vspace{-10pt}
\end{figure}

To make full-iteration simulation tractable, we combine coarse-grained HBM
approximation with exact HBF acceleration. Only HBM is approximated:
scheduling transfers per 4\,KiB chunk and pseudo-channel rather than per
32-byte burst runs an all-HBM forward pass 24.8$\times$ faster
(134.7$\times$ for sequential reads), while keeping batch completion times
within 0.04\%.

For HBF, the bottleneck is reserving roughly a dozen shared resources per
page, as in SimpleSSD~\cite{jung_simplessd}; here we reduce per-reservation
overhead without sacrificing accuracy. A first-fit calendar organizes each
resource's idle intervals into 64-interval leaves labeled with their maximum
duration, so searches immediately bypass leaves with insufficient space.
Multi-page reads are batched into groups of one page per plane to share
stage-time calculations, reuse calendar lookup states across groups, and
compute stride addresses from the first page, while the FTL updates free
space only for planes whose state changes. Page grouping speeds up a
Llama-3.1-70B batch (7.3M page reads) by 1.48$\times$, and lazy FTL
recounting accelerates ten write batches (3.2M page programs) by
1.55$\times$. Every page still reserves all resources in strict order
without pruning or rounding idle intervals, yielding bit-identical
completion times across all 82,449 transactions in the batch.

Wall-clock time then tracks flash page reads at roughly $2.7\times10^{5}$
reads/s per core (Figure~\ref{fig:validation}b); all 405 configurations in
Section~\ref{sec:cases} took 80.0 core-hours on an Apple M2 Ultra and an M5 Pro.

\vspace{-6pt}
\subsection{Calibration and Validation}
\label{sec:validation}

Three components are calibrated against hardware
(Figure~\ref{fig:validation}a). The scheduler is checked against vLLM~0.26's:
both receive the same requests and the same measured batch times, and over
282 batches in ten test cases with staggered arrivals, shared prefixes,
prefix eviction, and preemption, they form the same batches with the same
token ranges, prefix hits, and completion times. The operator model is
fitted to kernel measurements: on an A100 and an H200 NVL we timed the Marlin
GEMMs and FlashAttention-2 paged attention that vLLM uses with 8-bit
weights, inside whole layers under CUDA graphs, over 525 shapes per GPU
spanning batch, query, and context sizes, and fitted each operator's fixed
overhead, compute rate, and memory rate by least squares.
We then predict 64 whole-layer shapes that never entered the fit: the mean
absolute error is 5.8\% on the A100 and 5.0\% on the H200, and the largest,
41.8\% on the A100, occurs in Llama-3.1-70B layers whose batch mixes long
and short contexts. Finally we run complete requests on each GPU, from the
embedding to the output head, and compare each batch's time: predicted and
measured batch times differ by
10.2\% on average on the A100 (Llama-3.1-8B, 46 batches) and by 5.8\% on
the H200 (Llama-3.1-8B and Llama-3.1-70B, 86 batches); for Llama-3.1-70B on
the H200, the time to first token differs by 9.3--13.6\% and TPOT by
4.6--7.0\%. The host DRAM path is fitted to measured pinned-memory copies
between an H200 NVL and its host (issue time, bandwidth per direction,
fixed latency).

HBM4 and HBF cannot be calibrated yet. Their interfaces follow the
specifications; HBM's access latency and bandwidth efficiency and HBF's
internal timing (4\,$\mu$s page read, 75\,$\mu$s program) and parallelism
are assumptions users can replace. Every default parameter
records its source, published, from the literature, or assumed; 18 of the
30 parameter groups are assumptions. The engine checks its own accounting on every run (programs
by cause must sum to the physical programs; valid, invalid, and free pages
to the capacity), the test suite compares HBM scheduling, HBF reads and
writes, and GC against reference models, and every result records the
hashes of its inputs, configuration, and binary.

\section{Case Studies}
\label{sec:cases}

We use \hbfsim to answer the questions of Section~\ref{sec:questions} with two models and three workloads.

\vspace{-6pt}
\subsection{Setup}
\label{sec:setup}

\textbf{System.} We model one accelerator package with eight memory stacks,
each an HBM4 stack or an HBF stack (Table~\ref{tab:devices}); we write
2~HBM + 6~HBF for a package of two HBM and six HBF stacks (2:6 in the
figures, 2H6F in Table~\ref{tab:lifetime}). We compare 1~HBM + 7~HBF,
2~HBM + 6~HBF, 4~HBM + 4~HBF, and 6~HBM + 2~HBF, whose HBM holds 43.4,
91.9, 188.9, and 285.9\,GiB of data. The HBF stacks follow OCP speed grade~2, with 16
channels, one die per channel, 16 planes per die, and blocks of 256 pages of
4\,KiB. Like an SSD, each HBF stack has a device DRAM of 0.1\% of its flash
that serves as write buffer and read cache. Unless noted otherwise, the host
FTL uses page-level mapping and caches translation pages in 0.1\% of the HBF
capacity, reserved in HBM. We serve Llama-3.1-70B~\cite{dubey_llama3} with
8-bit weights (67.7\,GiB) and a BF16 KV cache (320\,KiB per token), and
Qwen3-235B-A22B with FP8 weights (223.0\,GiB for the 128 experts of each of
its 94 MoE layers) and a BF16 KV cache (188\,KiB per token); the figures
label them 70B and 235B. Public traces carry no expert routes, so the 235B
model routes every token to a uniformly random top-8 of its 128 experts,
seeded by the token's prefix; at 32 sequences a decode step reads
197.9\,GiB of the 223\,GiB of weights. GPU compute follows the H200 NVL
operator profile (Section~\ref{sec:validation}), which covers batches of up
to 16; for larger batches we keep the per-FLOP and per-byte costs measured
at 16.

\begin{table}[t]
    \caption{Memory devices of capacity tiers.}
    \label{tab:devices}
    \vspace{-8pt}
    \centering\footnotesize
    \setlength{\tabcolsep}{3pt}
    \begin{tabular}{@{}lll@{}}
        \toprule
        Device & Bandwidth & Latency and other parameters \\
        \midrule
        HBM4 stack, 48\,GiB & 2.0\,TB/s & 42\,ns read \\
        HBF stack, 512\,GiB & 1.5\,TB/s & 4\,$\mu$s read, 75\,$\mu$s program; \\
                            &           & device DRAM of 0.1\% of flash \\
        Host DRAM, PCIe~5.0 & 54\,GB/s & 9.3\,$\mu$s per transfer (measured) \\
        CXL SSD & 36\,GB/s link & 40\,$\mu$s NAND read, 6.6\,GB/s NAND; \\
                &               & DRAM cache of 5\% of flash, FIFO \\
        NVMe SSD & 16\,GB/s link & same NAND as the CXL SSD \\
        \bottomrule
    \end{tabular}
    \vspace{-8pt}
\end{table}

\begin{table}[t]
    \caption{Workloads at the representative decode step: prompt and output
    tokens per request, mean context of the running sequences (Ctx), KV read
    per step (KV, GiB), and working set of weights and KV (WS, GiB).}
    \label{tab:workloads}
    \vspace{-8pt}
    \centering\footnotesize
    \setlength{\tabcolsep}{3.2pt}
    \begin{tabular}{@{}lrrrrrrr@{}}
\toprule
& & & & \multicolumn{2}{c}{70B} & \multicolumn{2}{c}{235B} \\
\cmidrule(lr){5-6}\cmidrule(l){7-8}
Workload & INPUT & OUTPUT & Ctx & KV & WS & KV & WS \\
\midrule
Long OUTPUT & 2K & 6K & 5.2K & 51.2 & 118.9 & 30.1 & 253.0 \\
Balanced & 4K & 4K & 6.3K & 61.1 & 128.8 & 35.9 & 258.9 \\
Long INPUT & 6K & 2K & 7.3K & 70.9 & 138.6 & 41.7 & 264.7 \\
\bottomrule
\end{tabular}

    \vspace{-8pt}
\end{table}

\textbf{Workloads.} A decode step reads the KV of every running sequence,
so what distinguishes workloads is how much context the sequences hold and
how that context was produced. We characterize a workload by its
input:output length ratio and compare three (Table~\ref{tab:workloads}):
long OUTPUT (2K prompt tokens, 6K output tokens), balanced (4K and 4K), and
long INPUT (6K and 2K), 8K tokens per request in each. The public traces of
Mooncake~\cite{qin_mooncake} and Qwen~\cite{wang_kvcache_wild} are
input-heavy (at most 29\% of their requests have more than 62.5\% of their
tokens in the output), so no output-heavy workload can be drawn from them;
their median request holds 5.5K--7.3K tokens. Prompts
are private, so no KV is shared. We replay each workload through the
scheduler with 32 sequences per batch and a prefix cache that retains 1M
tokens (320\,GiB of KV for Llama-3.1-70B, 188\,GiB for Qwen3-235B-A22B).
The first batch's requests arrive spread over one request's lifetime, so in
the steady state the running sequences are at every stage of their
lifetimes. Among the steps in which every
batch slot is busy while requests wait and the prefix cache is full, we take
the one whose running sequences hold the median context and install its
state as replayed, each KV block at its replayed address. We then measure a
short window instead of a full replay: each sequence prefills a private
16-token suffix, which resumes it as a running request and yields its first
token, and decodes seven more; TPOT is the mean time of these seven decode
steps. The three
workloads hold 5.2K, 6.3K, and 7.3K tokens of context on average, so a step
reads 51--71\,GiB of KV for Llama-3.1-70B and 30--42\,GiB for
Qwen3-235B-A22B, 40\% more under long INPUT than under long OUTPUT, and
writes one token of KV per sequence.

\vspace{-6pt}
\subsection{HBF Versus Other Capacity Tiers}
\label{sec:tiers}

\begin{figure}[t]
    \centering
    \includegraphics[width=\columnwidth]{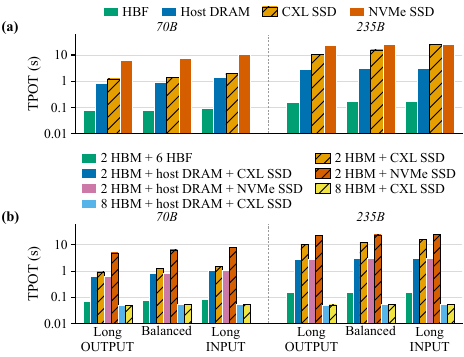}
    \vspace{-8pt}
    \caption{Mean TPOT at the decode step of each workload (logarithmic
    axes). (a) HBF against the other capacity tiers at 2~HBM + 6~HBF, peer
    mode with the weights placed first in HBM. (b) The tiers as systems with
    eight stack sites: 2~HBM + 6~HBF, or 2 or 8 HBM with the HBF's capacity
    off package as a CXL SSD or an NVMe SSD, alone or behind 256\,GiB of host
    DRAM; KV placed first in HBM.}
    \Description{Two rows of grouped bars for three workloads of Llama-3.1-70B
    and Qwen3-235B-A22B. Top: HBF takes 75 to 169 milliseconds per token,
    host DRAM 0.8 to 3.2 seconds, the CXL SSD 1.2 to 26 seconds, and the
    NVMe SSD 6.5 to 26 seconds. Bottom: 2 HBM + 6 HBF takes 69 to 160
    milliseconds, 2 HBM with host DRAM in front of either SSD 0.6 to 3.1
    seconds, 2 HBM with a CXL SSD alone 0.9 to 16 seconds, 2 HBM with an
    NVMe SSD alone 4.9 to 25 seconds, and 8 HBM with either tier 49 to 55
    milliseconds.}
    \label{fig:tiers}
    \vspace{-10pt}
\end{figure}

At 2~HBM + 6~HBF with weights first, the capacity tier serves 38--65\,GiB
of running KV per step for Llama-3.1-70B and 145--157\,GiB for
Qwen3-235B-A22B, whose 223\,GiB of weights exceed HBM: 115\,GiB of weights
and all 30--42\,GiB of running KV. We keep this placement and replace the six HBF stacks with host DRAM,
a CXL SSD, or an NVMe SSD of the same capacity (Figure~\ref{fig:tiers}a).
HBF decodes 11.0--14.4$\times$ faster than host DRAM, 16.1--21.3$\times$
faster than the CXL SSD, and 86.6--115.0$\times$ faster than the NVMe SSD
for Llama-3.1-70B, and 18.6--19.0$\times$, 66.8--154.1$\times$, and
150.3--154.1$\times$ faster for Qwen3-235B-A22B. The step time is roughly
the tier read divided by the tier's bandwidth: six on-package HBF stacks
serve 157\,GiB in 0.17\,s, PCIe-attached host DRAM in 3.2\,s, and the
NVMe SSD's NAND at 6.6\,GB/s in 26\,s. The CXL SSD's device DRAM (5\% of
its flash, 153.6\,GiB) holds the spill of Llama-3.1-70B and of the first
two Qwen3-235B-A22B workloads, so the CXL SSD stays within 1.5$\times$ of host
DRAM there; the 157\,GiB spill of long INPUT exceeds it and the CXL SSD is
as slow as the NVMe SSD (26.0\,s). We keep all 32 sequences in every system, so each tier serves the
same spill; a real system would shrink the batch.

An off-package tier occupies none of the package's eight stack sites,
which can then hold HBM. Figure~\ref{fig:tiers}b therefore compares seven systems on eight
sites, all with KV first: 2~HBM + 6~HBF; 2~HBM with the HBF's capacity off
package as a CXL or NVMe SSD, alone or behind 256\,GiB of host DRAM (one
accelerator's share of a 2\,TiB server), leaving six sites empty; and 8~HBM
with the CXL SSD, alone or behind host DRAM. With two HBM stacks, the
weights that do not fit in HBM come from the tier in every step:
2~HBM + 6~HBF is 9.3--12.5$\times$ faster than 2~HBM + host DRAM + CXL SSD
for Llama-3.1-70B and 19.1--19.5$\times$ for Qwen3-235B-A22B. Host DRAM
holds the first 256\,GiB of the tier, which covers both models' weights, so
the SSD behind it is never read and the NVMe and CXL variants coincide. The
SSDs alone are 13.5--18.3$\times$ (CXL) and 70.8--98.0$\times$ (NVMe) slower
than 2~HBM + 6~HBF for Llama-3.1-70B, and 66.9--99.3$\times$ and
154.3--157.2$\times$ for Qwen3-235B-A22B. With eight HBM stacks
(384\,GiB), both working sets fit, no step reads the off-package tier, and
TPOT is 49--55\,ms: 1.4--1.5$\times$ below 2~HBM + 6~HBF for
Llama-3.1-70B and 2.9$\times$ below it for Qwen3-235B-A22B. An off-package tier thus wins only
when the sites it frees go to HBM and the step fits in that HBM; then a
cheap tier serves as well as a fast one.

\takeaway{HBF is most valuable under memory-capacity pressure.}{It offers little benefit when the working set fits in HBM, but substantially outperforms offloading to CPU memory, CXL SSDs, and NVMe SSDs once spilling occurs, for example when Qwen3-235B-A22B serves 32 sequences whose 253--265\,GiB working set exceeds the 188.9\,GiB of HBM at 4~HBM + 4~HBF. Under a fixed stack budget, however, adding more HBM can hurt performance if the resulting loss of HBF bandwidth outweighs the reduction in spill traffic.}

\vspace{-6pt}
\subsection{How Much HBM, and What Should It Hold?}
\label{sec:hbm}

\begin{figure*}[t]
    \centering
    \includegraphics[width=\textwidth]{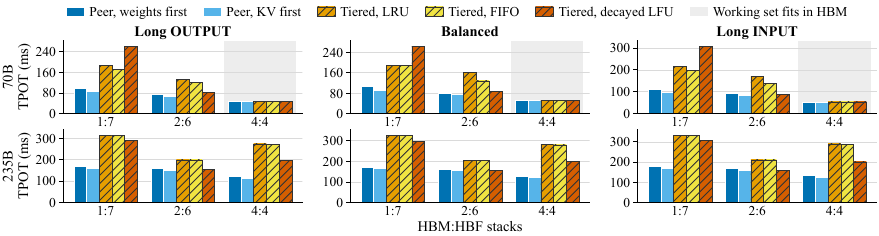}
    \vspace{-8pt}
    \caption{Mean TPOT across HBM:HBF stack splits. Peer mode gives HBM
    first to the weights or to the running sequences' KV; tiered mode
    replaces by LRU, FIFO, or decayed LFU and writes back on eviction.}
    \Description{Six bar charts, one per model and workload, with the
    HBM:HBF split (1:7, 2:6, 4:4) on the horizontal axis and TPOT in
    milliseconds on the vertical axis. For Llama-3.1-70B, the working set
    fits at 4:4, where every bar is 49 to 55 milliseconds except weights
    first under long OUTPUT at 75; at 2:6, KV first takes 69 to 83, weights
    first 75 to 95, tiered LRU 134 to 172, FIFO 122 to 138, and decayed LFU
    84 to 90; at 1:7, KV first takes 89 to 98, weights first 100 to 112,
    LRU 187 to 216, FIFO 171 to 197, and decayed LFU 261 to 308. For
    Qwen3-235B-A22B, the working set fits at none of the three splits; KV
    first falls from 163 to 170 at 1:7 and 152 to 160 at 2:6 to 115 to 128
    at 4:4, tiered LRU and FIFO take 315 to 333 at 1:7, 200 to 211 at 2:6,
    and 274 to 291 at 4:4, and decayed LFU 292 to 307, 156 to 161, and 196
    to 201.}
    \label{fig:hbm}
    \vspace{-10pt}
\end{figure*}

\textbf{Capacity and placement.} We vary the split between HBM and HBF
(Figure~\ref{fig:hbm}). In peer mode, HBM
should give priority to the running sequences' KV. The working set of Llama-3.1-70B
(119--139\,GiB) fits at 4~HBM + 4~HBF, where KV first and tiered
mode reach the compute-bound 49--54\,ms. At 2~HBM + 6~HBF, KV first takes
69--83\,ms and weights first 75--95\,ms, 4--14\% more: it fills the HBM
left after the weights with KV blocks in address order, retained prefixes
included, and leaves 38--65\,GiB of running KV in HBF per step, whereas KV
first reads only weights from HBF (29--49\,GiB). At 1~HBM + 7~HBF even KV first reads 8--28\,GiB
of running KV per step from HBF (89--98\,ms). Under long OUTPUT, weights
first still takes 75\,ms at 4~HBM + 4~HBF, where the working set fits,
because the added HBM fills with retained prefixes while 31\,GiB of
running KV per step still come from HBF. The working set of
Qwen3-235B-A22B (253--265\,GiB) exceeds HBM at all three splits. Its
decode step is dominated by the 198\,GiB of expert weights it reads, so KV
first falls only gradually, from 163--170\,ms at 1~HBM + 7~HBF and
152--160\,ms at 2~HBM + 6~HBF to 115--128\,ms at 4~HBM + 4~HBF, and
weights first is 4--7\% behind. At 1~HBM + 7~HBF the step takes only 7\% longer than
at 2~HBM + 6~HBF: seven HBF stacks serve the experts. The three workloads differ little: their contexts differ by 1K
tokens, and TPOT follows (69, 76, and 83\,ms with KV first at
2~HBM + 6~HBF): a step's cost depends on how much context it reads, not
on how it was produced.

\textbf{Tiered mode.} Tiered mode matches peer mode with KV first when the
working set fits in HBM and, with LRU or FIFO, is 1.7--2.2$\times$ slower
for Llama-3.1-70B and 1.3--2.4$\times$ for Qwen3-235B-A22B when it does
not. Each step reads the working set in the same order, a cyclic pattern larger
than HBM, so LRU evicts almost every 4\,MiB line before it is reused and the
step rereads the whole working set from HBF, 117--137\,GiB for Llama-3.1-70B
and 231--243\,GiB for Qwen3-235B-A22B, instead of only the spill. This is
why 4~HBM + 4~HBF is slower than 2~HBM + 6~HBF for Qwen3-235B-A22B in tiered
mode (275--291 against 201--211\,ms): four HBF stacks serve the same
reread instead of six. At 1~HBM + 7~HBF the single HBM stack, which also
absorbs every refill as a write, becomes the limit at 1.4--1.6\,TB/s of
its 2.0.

\textbf{Replacement.} No policy wins everywhere. FIFO rereads the same bytes
as LRU in another order: within 1\% of LRU for Qwen3-235B-A22B and
1--20\% faster for Llama-3.1-70B. Decayed LFU keeps the lines read most often: for
Llama-3.1-70B at 2~HBM + 6~HBF it keeps the running KV in HBM and refills
the weights every step, the division of KV first, and takes 84--90\,ms,
37--48\% less than LRU. At 1~HBM + 7~HBF, where the running KV
(51--71\,GiB) no longer fits, the same policy is the slowest, 261--308\,ms,
1.4$\times$ LRU: it evicts its newest lines first. For Qwen3-235B-A22B it
cuts LRU's time by 7--10\% at 1~HBM + 7~HBF and 22--31\% at the next two
splits but stays 1.0--1.8$\times$ above KV first: a newly filled line
starts with a count of one, so the expert weights, which every step reads
once, keep being evicted and refilled.

\begin{figure}[t]
    \centering
    \includegraphics[width=\columnwidth]{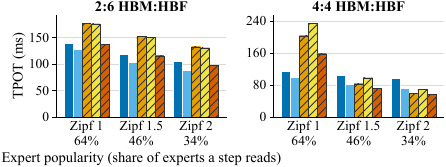}
    \vspace{-8pt}
    \caption{Qwen3-235B-A22B under skewed expert popularity (balanced
    workload; legend as in Figure~\ref{fig:hbm}): a token's top-8 experts
    are drawn with weight $(r+1)^{-\alpha}$ over a seeded per-layer rank
    $r$. Tick labels: share of the experts a step reads.}
    \Description{Two bar charts, for the 2:6 and 4:4 splits, with Zipf
    exponents 1, 1.5, and 2 on the horizontal axis (64, 46, and 34 percent
    of the experts read per step) and TPOT in milliseconds on the vertical
    axis. At 2:6, KV first takes 127, 104, and 88 milliseconds, decayed LFU
    137, 116, and 97, LRU 177, 152, and 133. At 4:4, KV first takes 100, 84,
    and 72, decayed LFU 159, 72, and 56, LRU 203, 83, and 59, and FIFO 236,
    98, and 69.}
    \label{fig:skew}
    \vspace{-10pt}
\end{figure}

\textbf{Hot experts.} The uniform router is the worst case for a cache.
We therefore rerun
Qwen3-235B-A22B with a Zipf router of exponent $\alpha$
(Figure~\ref{fig:skew}), so that a step reads 64\%, 46\%, and 34\% of the
experts at $\alpha$ = 1, 1.5, and 2 instead of 88\%.
At 4~HBM + 4~HBF, where $\alpha$ = 1.5 leaves 135--147\,GiB in 188.9, decayed
LFU takes 0.83--0.87 of the KV-first time and LRU 0.89--1.00; at
$\alpha$ = 2 they take 0.78 and 0.81--0.83: the cache keeps every layer's
hot experts, whereas the address-ordered split keeps whole early layers
and misses those of the late ones.
At 2~HBM + 6~HBF the split still wins at every exponent: decayed LFU
stays 1.07--1.10$\times$ above KV first, because the 26 whole layers in
HBM save their cold experts too while the cache spends its remainder on
cold lines that rarely recur, and LRU and FIFO 1.4--1.5$\times$ above.

\textbf{When each mode wins.} One quantity separates the two modes in
Figures~\ref{fig:hbm} and~\ref{fig:skew}: the set a step reads against the
HBM capacity. Peer mode with KV first wins whenever that set exceeds HBM, and the
margin must be real: at $\alpha$ = 1 and 4~HBM + 4~HBF the 180\,GiB set,
144\,GiB of weights and 36 of KV, nominally fits in 188.9, yet FIFO rereads
193\,GiB per step, the set in 4\,MiB lines, and LRU 161. Tiered mode wins only when the set fits with
room to spare and its reuse is spread across layers, which an
address-ordered split cannot follow; where the set fits outright, the two
modes are equal.

\takeaway{Under the evaluated workloads and 4 MiB cache-line configuration, the tested replacement policies incur substantial refill traffic when the active working set exceeds HBM capacity.}{None of the evaluated policies matches peer mode with KV first once the set a step reads exceeds HBM, and they beat it only when that set fits with room to spare, because none captures the distinct access patterns, mutability, and lifetimes of inference data. Effective HBF management therefore requires inference-aware placement and migration rather than generic cache replacement.
}

\begin{figure}[t]
    \centering
    \includegraphics[width=\columnwidth]{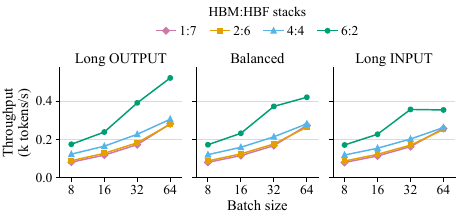}
    \vspace{-8pt}
    \caption{Decode throughput of Qwen3-235B-A22B against the batch size,
    peer mode with KV first.}
    \Description{Three line charts, one per workload, with batch sizes 8,
    16, 32, and 64 on the horizontal axis and decode throughput in
    thousands of tokens per second on the vertical axis, one line per split.
    Throughput grows from 80 to 174 tokens per second at batch 8 to 251 to
    520 at batch 64. The 6:2 line is the highest at every batch and rises to
    520 under long OUTPUT, 420 under balanced, and 353 under long INPUT at
    batch 64; the 1:7, 2:6, and 4:4 lines are within 10\% of one another at
    batch 64, and the 1:7 line, lowest at batch 8, overtakes 2:6 at batch
    64.}
    \label{fig:batch}
    \vspace{-10pt}
\end{figure}

\textbf{Batch size.} We repeat the Qwen3-235B-A22B study in peer mode with
KV first at batches of 8, 16, 32, and 64, each from its own replay
(Figure~\ref{fig:batch}). Throughput grows 2.1--3.0$\times$ from batch 8
to 64 at 6~HBM + 2~HBF and 2.9--3.5$\times$ at 2~HBM + 6~HBF and
1~HBM + 7~HBF, while TPOT grows from 37\,ms to 67--117\,ms and from
74--84\,ms to 186--205\,ms. The working set grows with the batch, to 283,
295, and 306\,GiB at batch 64; only long OUTPUT still fits in the 285.9\,GiB of 6~HBM + 2~HBF, its
520 tokens/s compare with 420 and 353 for the other two, and tiered mode
with LRU, 67\,ms where the working set fits, takes 503 and 672\,ms once
it exceeds HBM by 3\% and 7\%. With one to four HBM stacks the expert
weights come from HBF in any case; the second pair of HBM stacks is worth
38\% of throughput at batch 8 but 9\% at batch 64.

\textbf{Write path.} A decode step writes too little KV to compare ways of
writing it, so we also serve 112, 56, and 38 requests of the three
workloads from an empty KV state at 2~HBM + 6~HBF (outputs capped at
8 tokens), which writes 46--47\,GiB of KV per workload. Writing the KV that does not fit in HBM straight to HBF, writing
all KV into HBM and to HBF when its line is evicted, and draining it to HBF
in the background give the same mean request time within 1.5\%, because
prefill is compute-bound and each HBF stack buffers writes in its device
DRAM. The background drain, however, writes 70--71\,GiB to flash,
1.2--1.6$\times$ as much as write-back on eviction (45--60\,GiB). In decode,
KV first never programs an HBF page, while tiered mode with LRU programs
61,000--65,000 pages per step for Qwen3-235B-A22B as it evicts dirty KV.

\takeaway{KV data should be prioritized in HBM and moved to HBF only under capacity pressure.} {KV is continuously appended during decoding, making its placement a primary determinant of HBF write traffic. Allocating KV to HBF only when it must leave HBM incurs no performance loss in our experiments, whereas proactive migration creates avoidable writes for KV entries that may be freed shortly.}

\vspace{-6pt}
\subsection{Flash Translation and Endurance}
\label{sec:endurance}

\begin{figure}[t]
    \centering
    \includegraphics[width=\columnwidth]{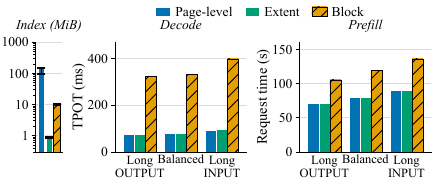}
    \vspace{-8pt}
    \caption{Page-level, extent, and block mapping for Llama-3.1-70B at
    2~HBM + 6~HBF, weights first: the index kept for the pages a decode
    step touches (bar: largest workload; whisker: smallest), decode TPOT, and
    request time in the prefill windows of Section~\ref{sec:hbm}.}
    \Description{Three panels. Left: index memory on a logarithmic axis,
    page-level 97 to 152 mebibytes, extent 0.9 mebibytes, block 10
    mebibytes. Middle: decode TPOT per workload; page-level and extent bars
    are equal at 75 to 95 milliseconds, block bars 4.2 to 4.3 times higher.
    Right: prefill request time; page-level and extent equal at 71 to 90
    seconds, block 1.5 times higher.}
    \label{fig:mapping}
    \vspace{-10pt}
\end{figure}

In decode, HBF mostly serves reads, so we ask what the FTL costs on the
read path and how fast writes wear HBF.

\textbf{Address translation.} We compare the three mappings of
Section~\ref{sec:device} at 2~HBM + 6~HBF with weights first, so that the
running KV is read from HBF (Figure~\ref{fig:mapping}). A decode step writes
almost nothing, so we also run the prefill windows of Section~\ref{sec:hbm}
(46--47\,GiB of new KV each). All three receive the same 3.1\,GiB of HBM, but
only page-level mapping uses a sizable part of it: 97--152\,MiB of
translation pages for the pages a decode step touches, against 0.9\,MiB for
the extent index and 10\,MiB for the block index. Extent mapping stays within
0.5\% of page-level mapping in decode and in prefill: KV is written and read
in long runs, so one index entry serves many pages. Block-level mapping is
4.2--4.3$\times$ slower in decode (324--399 against 75--95\,ms), because a
block lies on one plane, which reads its 256 pages one after another in about
1.0\,ms, whereas the other two mappings spread the same pages over all planes;
prefill, which spends more of its time in compute, hides part of this and is
1.5$\times$ slower.

\takeaway{HBF mapping should preserve internal parallelism rather than maximize address granularity.}{Decode reads weights and KV in long sequential runs, so one translation serves many pages. An extent mapping that stripes each run across all planes matches a page-level table at a fraction of its memory, whereas a block-level mapping that keeps each block on a single plane serializes its page reads and slows decoding considerably.
}

\begin{table}[t]
        \caption{Estimated HBF lifetime in months at the published H200 peak
    load of Llama-3.3-70B (5,000-token prompts, 500-token outputs), with 100,000 P/E cycles.}
    \label{tab:lifetime}
    \vspace{-8pt}
    \centering\footnotesize
    \setlength{\tabcolsep}{3pt}
    \begin{tabular*}{\columnwidth}{@{\extracolsep{\fill}}lrrrr@{}}
\toprule
% & \multicolumn{4}{c}{Page-level mapping} \\
% \cmidrule{2-5}
Workload & 1H7F & 2H6F & 4H4F & 6H2F \\
\midrule
\multicolumn{5}{@{}l}{Llama-3.1-70B} \\
\quad Long OUTPUT & 83.0 & 71.2 & 47.4 & 23.7 \\
\quad Balanced & 83.0 & 75.9 & 50.6 & 25.3 \\
\quad Long INPUT & 83.0 & 71.1 & 54.2 & 27.1 \\
\multicolumn{5}{@{}l}{Qwen3-235B-A22B} \\
\quad Long OUTPUT & 141.4 & 121.2 & 80.8 & 40.4 \\
\quad Balanced & 141.4 & 121.2 & 80.8 & 43.1 \\
\quad Long INPUT & 141.4 & 121.2 & 80.8 & 46.2 \\
\bottomrule
\end{tabular*}

    \vspace{-8pt}
\end{table}

\textbf{Lifetime.} A decode step writes one token of KV per sequence, so
wear comes from new KV written into flash, and we estimate lifetime at a
published peak load. TensorRT-LLM reaches 489 output tokens/s per H200 for
Llama-3.3-70B, which has the architecture of Llama-3.1-70B, with 5,000-token
prompts and 500-token outputs~\cite{nvidia_trtllm_perf}, 1.64\,GiB/s of new
KV per GPU. The published table has no Qwen3-235B-A22B, so we apply the same
token rate to it; with 188\,KiB of KV per token against 320\,KiB, it writes
41\% less.
\hbfsim measures the share of this KV that reaches HBF with weights first
(87.5--100\%) and replays each workload's KV writes and releases (about
1\,TiB, twice the stack's capacity) through one HBF stack with garbage
collection, which gives the flash bytes erased per written byte in steady
state, the quantity that consumes P/E cycles. With either mapping this is 1.0 within 0.1\%: KV
is appended and freed in bulk, so one block with no valid page left is
erased per MiB written. Block-level mapping programs 1.5--3.5\% fewer
pages, because pages that wait in the device DRAM for their block's
unwritten prefix are dropped when a sequence ends, but their block is still
erased. Assuming 100,000 P/E cycles, as for the SLC NAND used in HBF
designs~\cite{wang_flashaccel}, Table~\ref{tab:lifetime} gives the
resulting lifetime: 24--83 months for Llama-3.1-70B and 40--141 months for
Qwen3-235B-A22B, in proportion to the HBF stacks that share the writes and
the same with either mapping: how long HBF lasts is set by how much KV
reaches flash.

\takeaway{HBF endurance is a first-order system constraint.}{Under baseline inference workloads, KV spill generates enough writes to make projected HBF lifetime a major challenge even when performance is acceptable. Practical HBF systems therefore require write reduction and endurance-aware management.
}

\vspace{-6pt}
\subsection{Which HBF Parameters Matter?}
\label{sec:parameters}

\begin{figure}[t]
    \centering
    \includegraphics[width=\columnwidth]{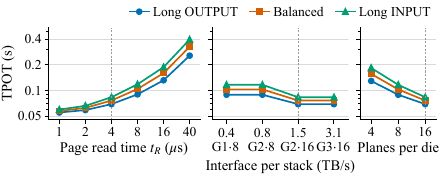}
    \vspace{-8pt}
    \caption{TPOT of Llama-3.1-70B at 2~HBM + 6~HBF, KV first, when one HBF
    parameter changes (logarithmic axis). Dashed: default setting. Interface points give the OCP speed grade and
    channels per stack; 8-channel stacks have half the planes.}
    \Description{Three line charts, one line per workload. TPOT grows almost
    in proportion to the page read time above 4 microseconds and keeps
    falling to 1 microsecond. Speed grade 3 on 16 channels equals speed
    grade 2; the two 8-channel points are equally slower. Halving the planes
    per die from 16 to 8 and to 4 raises TPOT like doubling and quadrupling
    the read time.}
    \label{fig:parameters}
    \vspace{-10pt}
\end{figure}

The HBF specification is still a draft, so we ask which device parameters
the results depend on, taking Llama-3.1-70B on the three workloads at
2~HBM + 6~HBF with KV first, where a step reads 29--49\,GiB of weights
from HBF (69--83\,ms per token), and changing one parameter at a time
(Figure~\ref{fig:parameters}).

Only the rate at which the flash array senses pages matters. Doubling the
page read time to 8\,$\mu$s raises TPOT by 30--42\%, 16\,$\mu$s by
91--126\%, and the 40\,$\mu$s of conventional NAND by 3.7--4.8$\times$;
halving the planes per die at equal capacity has the same effect as
doubling the read time (29--40\% for 8 planes, 87--121\% for 4). The
interface does not matter: speed grade~3 doubles grade~2's 1.5\,TB/s and
changes TPOT by 0.1\%, and grades~1 and~2 on 8 channels are equally
29--40\% slower, only because their stacks have half the planes. A plane
senses one 4\,KiB page per read time, so a stack's 256 planes deliver
0.26\,TB/s at 4\,$\mu$s, a sixth of the interface. The write side is
invisible in decode, even without the device DRAM: with KV first a step
appends its KV in HBM and programs no HBF page, so program times from 37.5
to 600\,$\mu$s, write credits, and write timeout leave TPOT unchanged, and
the thermal throttling ceiling adds at most 0.2\%. For read-heavy decode, an
HBF design should therefore spend on planes and read time before interface
speed.

\section{Discussion}
\label{sec:discussion}

\textbf{Limitations.}
Our case studies reproduce one decoding step per workload, seven tokens per
running sequence, and one prefill window per
workload for the write path and the mapping comparison, without garbage
collection, so TPOT omits its interference with reads. We model
one accelerator; tensor or pipeline parallelism would give each a share of
the weights and KV. The
studies combine an H200 compute profile, extended beyond its measured
batch of 16, with HBM4 and HBF, a pairing that does not exist today, so
HBF timing and endurance remain assumptions. The lifetime estimates apply a published serving load
around the clock to the HBF write share and write amplification that
\hbfsim measures, and assume perfect wear leveling.

\textbf{Extending \hbfsim.}
New memory policies use the policy interface, new FTL schemes the host
mapping interface, new HBF generations the device configuration, and new
models, traces, and GPUs their descriptors, importers, or operator profiles.

\section{Related Work}
\label{sec:related}

\textbf{HBF and hybrid HBM--flash memory.}
H3~\cite{ha_h3} and SK hynix~\cite{skhynix_h3} evaluate hybrid HBM+HBF
organizations with projected device parameters. FlashAccel co-designs data
layout and storage management for HBF~\cite{wang_flashaccel}. Petrucci
et al.\ compare HBF with LPDDR analytically and identify writes as the
limit for offloading KV to flash~\cite{petrucci_hbf_all_you_need}.
\hbfsim lets such designs be compared on the same workloads; our endurance
results quantify the write problem they point out.

\textbf{Memory and storage simulators.}
Ramulator~2.0~\cite{luo_ramulator2} and DRAMsim3~\cite{li_dramsim3} model
DRAM at the command level. MQSim~\cite{mqsim} and FEMU~\cite{femu} model
SSDs behind an NVMe interface, with mapping, GC, and wear leveling inside the
device, and Cylon~\cite{cylon} emulates CXL SSDs with a DRAM cache in the
device. \hbfsim follows MQSim's approach to the flash organization and
Cylon's device class for the CXL SSD, but treats HBF as memory with a host-side FTL whose metadata
competes with the application in HBM.

\textbf{LLM serving simulators.}
Vidur~\cite{agrawal_vidur}, LLMServingSim~\cite{cho_llmservingsim},
Frontier~\cite{feng2026frontiercomprehensiveaccuratellm}, and
ASTRA-sim~\cite{won_astrasim3} simulate scheduling, computation, and
communication for LLM inference, LLMCompass~\cite{zhang_llmcompass} adds
operator-level modeling with mapping search, and GTSim~\cite{gtsim} and
Accel-Sim~\cite{khairy_accelsim,accelsim2_2026} model GPU microarchitecture. They represent memory by its capacity and
bandwidth; \hbfsim models the devices at the transaction level, flash
management included, under a validated scheduler.

% \textbf{Memory management for LLMs.}
% PagedAttention~\cite{kwon_pagedattention} and SGLang~\cite{zheng_sglang}
% manage KV memory within HBM. Mooncake~\cite{qin_mooncake},
% FlexGen~\cite{sheng_flexgen}, and ZeRO-Infinity~\cite{rajbhandari_zeroinfinity}
% offload KV or weights to host memory and SSDs, while H2O~\cite{zhang_h2o} and
% RetrievalAttention~\cite{liu_retrievalattention} reduce the KV that must be
% kept or read. \hbfsim can evaluate such techniques with HBF as the
% offloading target.
\vspace{-7pt}
\section{Conclusion}

HBF could give accelerators an order of magnitude more memory, but only if
systems use it in ways that suit flash. \hbfsim lets researchers test these
choices before hardware exists. Our case studies support five
takeaways. HBF is most valuable
under capacity pressure, where it keeps TPOT 11.0--19.0$\times$ lower than
host DRAM and 86.6--154.1$\times$ lower than an NVMe SSD. Conventional
cache replacement is a poor fit for HBM--HBF tiering: no policy matches
peer mode with KV first once the set a step reads exceeds HBM, and a cache
wins only when that set fits with room to spare. HBM should hold the
running sequences' KV and HBF the read-only weights; KV should move to HBF
only when it must leave HBM. The mapping should preserve the flash array's parallelism,
which an extent index does at a fraction of a page table's memory.
Endurance is a first-order constraint: at a published peak load of
Llama-3.1-70B, HBF lasts 2.0 to 6.9 years, in proportion to the stacks
that share the KV writes. We
release \hbfsim so that others can replace our assumptions with
measurements.
\vspace{-5pt}

% Start the template's balancing package while still in the first column.
\balance
\bibliographystyle{ACM-Reference-Format}
\bibliography{refs}

\end{document}